\documentclass[aps,prb,preprint,groupedaddress]{revtex4-1}

\usepackage{bm}
\usepackage{amsmath}
\usepackage{graphicx}

\newcommand{\Gru}{Gr{\"u}neisen }
\newcommand{\mode}{\omega_{n,\mathbf{k}}}
\newcommand{\mgp}{\gamma_{n,\mathbf{k}}}

\begin{document}


\title{Origins of anisotropic thermal expansion in flexible  materials}



\author{Carl P. Romao}
\affiliation{Department of Chemistry, University of Oxford, Inorganic Chemistry Laboratory, South Parks Road, Oxford OX1 3QR, UK}


\date{\today}

\begin{abstract}
A definition of the \Gru parameters for anisotropic materials is derived based on the response of phonon frequencies to uniaxial stress perturbations. This \Gru model relates the thermal expansion in a given direction ($\alpha_{ii}$) to one element of the elastic compliance tensor, which corresponds to the Young's modulus in that direction ($Y_{ii}$). The model is tested through \emph{ab initio} prediction of thermal expansion in zinc, graphite, and calcite using density functional perturbation theory, indicating that it could lead to increased accuracy for structurally complex systems. The direct dependence of $\alpha_{ii}$ on $Y_{ii}$ suggests that materials which are flexible along their principal axes but rigid in other directions will generally display both positive and negative thermal expansion.
\end{abstract}


\maketitle

\section{Introduction}

Materials which lack cubic symmetry will expand (or contract) at different rates in different directions in response to a change in temperature. Thermal expansion anisotropy has been the subject of considerable recent attention due to the discovery of flexible framework materials with unusually large positive or negative coefficients of thermal expansion (CTEs) along one or two crystal axes. \cite{goodwin2008colossal, goodwin2009thermal, nanthamathee2014contradistinct, cai2014giant, takenaka2017colossal, dove2016negative} However, anisotropy has a long history of  complicating fundamental understanding of the origins of thermal expansion, \cite{kopp1852expansion} and, owing to the thermal stress introduced in consolidated polycrystals, anisotropy can limit the practical uses of materials. \cite{cheng1998thermal,romao2016relationships}

In some cases the origin of thermal expansion anisotropy can be appreciated intuitively by inspection of the structure: the interatomic interactions in graphite are obviously stronger within the graphene layers than between them. In other cases the relationship is more subtle, \emph{e.g.}, temperature-induced displacive phase transitions in quartz and cristobalite introduce significant thermal expansion anisotropy while retaining the network topology.\citep{barron1982thermal,schmahl1992landau} The orthorhombic Sc$_2$W$_3$O$_{12}$ structure, which produces characteristically large anisotropy between axes with negative and positive CTEs, is isomorphic to the cubic aluminosilicate framework of garnet. \cite{romao2016relationships,evans1998negative} In the metal-organic wine-rack framework material MIL-53 replacement of an OH$^-$ anion by F$^-$ leaves the crystallographic symmetry unchanged but significantly modifies the thermal expansion anisotropy, changing the volumetric CTE ($\alpha_V$) from positive to negative. \citep{nanthamathee2014contradistinct}

The origins of thermal expansion in crystalline solids are commonly studied through a model originated by \Gru \cite{gruneisen1912theorie} which relates the contribution of a phonon to the thermal expansion to the volume derivative of its frequency. The \Gru approach is useful because changes in phonon frequencies as a function of volume can be measured using variable-pressure inelastic scattering techniques and calculated \emph{ab initio} using, for example, density-functional perturbation theory (DFPT), allowing explication of the mechanisms of thermal expansion. \citep{mittal2001origin, zwanziger2007phonon, zhou2008origin, peterson2010local, yoon2011negative, rimmer2014acoustic, rimmer2014framework} However, this model does not consider material anisotropy and an extension, incorporating coupling between elastic anisotropy and thermal expansion anisotropy, is required for non-cubic crystal families. 

The most notable such extension, based on replacing the volume perturbation by uniaxial strain perturbations, was developed by Barron and Munn \cite{barron1967analysis} following Gr{\"u}neisen's approach. \cite{gruneisen1924untersuchungen} Due to the experimental challenges involved in applying uniaxial strain to a sample, \citep{jones1996diamond} the anisotropic \Gru theory of Ref. \citenum{barron1967analysis} has until recently been used primarily to calculate directional \Gru parameters from experimental thermal expansion and heat capacity data,  \citep{bailey1970anisotropic, yates1975anisotropic, sears1979polytypism, barron1982thermal, huang2016temperature, romao2016relationships} and to identify the contributions of acoustic modes to thermal expansion anisotropy. \citep{ramachandran1972generalised} Therefore, until the fairly recent development of \emph{ab initio} methods which could calculate phonon band structures as a function of an arbitrary strain, the ability of the Barron--Munn model to predict anisotropic thermal expansion had been untested. \emph{Ab initio} prediction of thermal expansion anisotropy has shown results mixed between qualitative and quantitative levels of accuracy. \citep{fang2014ag, arnaud2016anisotropic, wang2016correlation, murshed2016thermal, liu2017anisotropic}

\begin{figure}[htbp]
 \includegraphics[width=10cm]{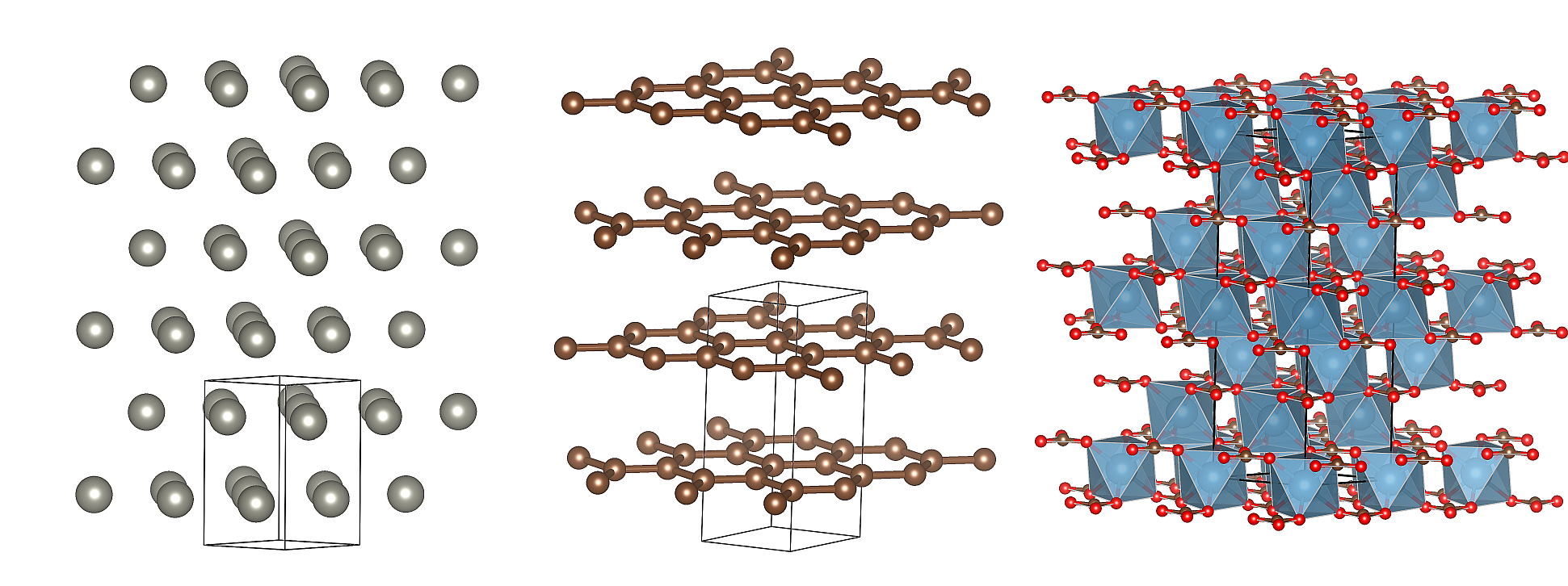}
 \caption{\label{cifs}
Crystal structures of materials with highly anisotropic thermal expansion and elastic properties used herein to test anisotropic \Gru models. \cite{icsd} From left to right: zinc, graphite, calcite. The $c$ axes are aligned vertically.}
\end{figure}

In order to understand and predict the behaviour of flexible  materials, defined here as those with some elastically compliant direction, we must understand how thermal expansion and elasticity are coupled. To further this goal, herein a \Gru model based on uniaxial stress perturbations is reported, which allows an explicit treatment of the coupling between \Gru parameters along different axes. The ability of the uniaxial stress model to predict axial CTEs is compared to that of the uniaxial strain model through DFPT calculations on several simple highly anisotropic materials (Fig. \ref{cifs}).

\section{Gr{\"u}neisen Models}

\subsection{The Isotropic \Gru Model}

To understand the place of anisotropy within the \Gru formalism, it is instructive to begin with a brief discussion of the original \Gru model for isotropic or cubic systems. The thermodynamic \Gru parameter ({$\gamma$}) is introduced through the identity
\begin{align}
\alpha_V = \frac{1}{V} \left( \frac{\partial V}{\partial{T}}\right)_P = \frac{1}{K_T}\left( \frac{\partial S}{\partial{V}}\right)_T = \frac{1}{V}\frac{\gamma C_V}{K_T},
\label{Tgru}
\end{align}
where the quantity $\gamma C_V$ represents a `phonon pressure', resulting from vibrational anharmonicity, which acts against the bulk modulus ($K_T$) to change the dimensions of the unit cell. Using the quasiharmonic approximation (QHA), the contribution of an individual phonon mode with frequency $\mode$ to the thermal expansion is determined through the mode \Gru parameter ($\mgp$), where
\begin{align}
\mgp = -\frac{V}{\mode}\left(\frac{\partial \mode}{\partial V}\right)_T.
\label{mgru}
\end{align}
Then, $\gamma$ and $\mgp$ are related by
\begin{align}
\gamma = \frac{\sum_{n,\mathbf{k}} \mgp C_{V,n,\mathbf{k}}}{\sum_{n,\mathbf{k}} C_{V,n,\mathbf{k}}}.
\label{grusum}
\end{align}
Differences between $\gamma$ as defined by Eq. (\ref{Tgru}) and $\gamma$ as defined by Eq. (\ref{grusum}) are due to anharmonic phonon-phonon interactions, and therefore are reduced with decreasing temperature. \citep{fultz2010vibrational} The exact validity of Eqs. (\ref{Tgru}--\ref{grusum}) also requires elastic isotropy of the lattice vectors and internal strain coordinates. \citep{hofmeister2002redefinition} When cubic symmetry is not present, the phonon frequencies do not depend only on the volume of the system, but also on the combination of strains required to reach a given volume from the equilibrium state.

\subsection{Uniaxial Strain Models}

Barron and Munn defined \Gru parameters for the response of a phonon to a (uniaxial) Lagrangian strain ($\eta_{ij}$) as: \citep{barron1967analysis}
\begin{align}
\hat{\gamma}_{ij,n,\mathbf{k}} = - \frac{1}{\mode} \left( \frac{ \partial \mode}{\partial \eta_{ij}} \right)_{\eta_{kl \neq ij}}.
\label{bmgamma}
\end{align}
Following averaging, by analogy to Eq. (\ref{grusum}) the directional thermal expansion is then constructed as: \citep{barron1967analysis}
\begin{align}
\alpha_{ij} = \left( \frac{\partial \eta_{ij}}{\partial T} \right)_{\bm{t}} = \frac {C_{\bm{\eta}}} {V} \sum_{kl} s_{ijkl} \hat{\gamma}_{ij} ,
\label{bmalpha}
\end{align}
where $s_{ijkl}$ are elements of the isothermal compliance tensor. 
 Note that the directional thermal expansion is defined here as a derivative under conditions of constant `thermodynamic tension' ($\bm{t}$), where
\begin{align}
t_{ij} = \left( \frac{\partial F}{\partial \eta_{ij}} \right)_{\eta_{kl \neq ij},T}.
\end{align}
Therefore, the perturbation in Eq. (\ref{bmgamma}) is uniaxial in terms of strain and thermodynamic tension, but not stress, since there are generally stresses in the directions perpendicular to $ij$ induced by the Poisson effect. These transverse stresses are accounted for in Eq. (\ref{bmalpha}) by linking the directional \Gru parameters through the cross-compliances, which assumes a mechanical coupling between the axial CTEs.

Choy \emph{et al.} \citep{choy1984thermal} treated Barron and Munn's definition of the \Gru parameter as arbitrary, and instead assumed an expression intermediate between Eq. (\ref{Tgru}) and Eq. (\ref{bmalpha}):
\begin{align}
\alpha_{ij} = \frac{1}{V}\frac{\tilde{\gamma}_{ij} C_{\bm{\eta}}}{3 K_T}.
\label{choyalpha}
\end{align}
However, this model is necessarily limited by its neglect of elastic anisotropy, and has been used sparingly for \emph{ab initio} prediction of thermal expansion. \citep{wdowik2011structural}



\subsection{Uniaxial Stress Model}

The derivation of a \Gru model based on uniaxial stress perturbations begins by considering the thermal expansion of a volume ($V$) under a constant stress ($\bm{\sigma}$). This stress is treated as a Cauchy stress, \emph{i.e.}, the volume of the stress-free reference state ($V_0$) is approximately equal to $V$. Accordingly, the conjugate infinitesimal strain ($\bm{e}$) is used, leading to the definition of thermal expansion used experimentally in the limit of small strains. Then, an arbitrary element of the thermal expansion tensor ($\bm{\alpha}$) is related to an uniaxial stress perturbation as
\begin{align}
\label{alphaij}
\alpha_{ij} &= \left( \frac{\partial e_{ij}}{\partial T} \right)_{\bm{\sigma}} = \left( \frac{\partial e_{ij}}{\partial \sigma_{ij}} \right)_{T, \sigma'} \left( \frac{\partial \sigma_{ij}}{\partial T} \right)_{e_{ij}, \sigma'}
 = s_{ijij} \left( \frac{\partial \sigma_{ij}}{\partial T} \right)_{e_{ij}, \sigma'},
\end{align}
where the subscript $\sigma'$ indicates that the elements of $\bm{\sigma}$ other than $\sigma_{ij}$ are kept constant. The relationship between $\sigma_{ij}$ and the free energy is then considered:
\begin{subequations}
\label{dFdeij}
\begin{align}
\left( \frac{\partial F}{\partial e_{ij}} \right)_{T, \sigma'} &= - \frac{V_0}{2} \left( \frac{\partial \left( \bm{\sigma}:\bm{e} \right)}{\partial e_{ij}} \right)_{T, \sigma'} \\
&= - V_0 \left(\sigma_{ij} + \frac{1}{2} \sum_{kl \neq ij} \sigma_{kl} \left( \frac{\partial e_{kl}}{\partial e_{ij}} \right)_{T, \sigma'} \right)
\end{align} 
\end{subequations} 
\begin{align}
\sigma_{ij} = -\frac{1}{V_0} \left( \frac{\partial F}{\partial e_{ij}} \right)_{T, \sigma'} +\frac{1}{2} \sum_{kl \neq ij} \sigma_{kl} \nu_{ijkl}.
\end{align}
 By substitution, 
\begin{align}
\alpha_{ij} = s_{ijij} \frac{\partial}{\partial T} \left(- \frac{1}{V_0} \left(\frac{\partial F}{\partial e_{ij}} \right)_{T, \sigma'} +\frac{1}{2} \sum_{kl \neq ij} \sigma_{kl} \nu_{ijkl} \right)_{e_{ij},\sigma'},
\end{align}
and, using the QHA,
\begin{align}
\alpha_{ij} = -s_{ijij} \left( \sum_{n,\mathbf{k}} \frac{\hbar}{V_0} \left( \frac{\partial \mode}{\partial e_{ij}} \right)_{T,\sigma'} \left( \frac{\partial}{\partial T} \frac{1}{e^{k_{\text{B}}T\mode}-1} \right)_{e_{ij},\sigma'} -\frac{1}{2} \sum_{kl \neq ij} \sigma_{kl} \frac{\partial \nu_{ijkl}}{\partial T} \right).
\end{align}

At this point $C_{\sigma'}$, the heat capacity under conditions of constant strain along $ij$ and constant stress along $kl \neq ij$, is introduced:
\begin{align}
C_{\sigma'} = T \left( \frac{\partial S}{\partial T} \right)_{e_{ij},\sigma'} =  \sum_{n,\mathbf{k}} C_{\sigma', n, \mathbf{k}} =  \sum_{n,\mathbf{k}} \hbar \mode \left( \frac{\partial}{\partial T} \frac{1}{e^{k_{\text{B}}T\mode}-1} \right)_{e_{ij},\sigma'}.
\end{align}
This heat capacity can be compared to $C_{\bm{e}}$ as follows:
\begin{subequations}
\begin{align}
\label{Cdef}
C_{\sigma'} &= T \left( \frac{\partial S}{\partial T} \right)_{\bm{e}} + T \sum_{kl \neq ij} \left( \frac{\partial S}{\partial e_{kl}}  \right)_{T, \sigma_{mn \neq kl}} \left( \frac{\partial e_{kl}}{\partial T}  \right)_{\bm{\sigma}} \\
&= C_{\bm{e}} + {T}{V_0} \sum_{kl \neq ij} \left( \left( \frac{\partial \sigma_{kl}}{\partial T}  \right)_{e_{kl}, \sigma_{mn \neq kl}} - \frac{1}{2} \sum_{mn \neq kl} \sigma_{mn} \frac{\partial \nu_{klmn}}{\partial T} \right) \alpha_{kl} \\
&= C_{\bm{e}} + {T}{V_0} \sum_{kl \neq ij} \left( \frac{\alpha_{kl}}{s_{klkl}} - \frac{1}{2} \sum_{mn \neq kl} \sigma_{mn} \frac{\partial \nu_{klmn}}{\partial T} \right) \alpha_{kl},
\end{align}
\end{subequations}
making use of Eqs. (\ref{alphaij}) and (\ref{dFdeij}). Then, the \Gru parameters are defined:
\begin{align}
\label{yunidef}
\check{\gamma}_{ij,n,\mathbf{k}} &= - \frac{1}{\mode} \left( \frac{\partial \mode}{\partial e_{ij}} \right)_{T,\sigma'} = - \frac{1}{s_{ijij}} \left( \frac{\partial ~\text{ln}~ \mode}{\partial \sigma_{ij}} \right)_{T,\sigma'} \\
\check{\gamma}_{ij} &= \frac{\sum_{n,\mathbf{k}} \check{\gamma}_{ij,n,\mathbf{k}} C_{\sigma',n,\mathbf{k}}}{\sum_{n,\mathbf{k}} C_{\sigma',n,\mathbf{k}}},
\end{align}
leading to the following expression for $\alpha_{ij}$
\begin{align}
\label{complex}
\alpha_{ij} = s_{ijij} \left( \check{\gamma}_{ij} \frac{C_{\sigma'}}{V_0} +\frac{1}{2} \sum_{kl \neq ij} \sigma_{kl} \frac{\partial \nu_{ijkl}}{\partial T} \right). 
\end{align}
By assuming that the external stress or the temperature derivatives of the transverse Poisson ratios are negligible, and that $C_{\sigma'} \approx C_{\bm{e}}$, the simplified expression
\begin{align}
\label{simple}
\alpha_{ij} = s_{ijij}  \check{\gamma}_{ij} \frac{C_{\bm{e}}}{V_0}
\end{align}
is obtained. For tetragonal and hexagonal crystal families, it is desirable to consider a biaxial stress perturbation along $a$ and $b$ in order to preserve phonon degeneracies. \citep{gan2016direct} Therefore, analogous areal versions of Eqs. (\ref{yunidef}), (\ref{complex}), and (\ref{simple}) are required:
\begin{align}
\check{\gamma}_{A,n,\mathbf{k}} &= - \left( \frac{\partial ~\text{ln}~\mode}{\partial ~\text{ln}~A} \right)_{T,\sigma_{cc}} \\
\alpha_{aa} &= \left(s_{aaaa}+s_{aabb}\right) \left( \check{\gamma}_{A} \frac{C_{\sigma_{cc}}}{V_0} +\frac{1}{2} \sigma_{cc} \frac{\partial \nu_{aacc}}{\partial T} \right)\\
\alpha_{aa} &= \left(s_{aaaa}+s_{aabb}\right)  \check{\gamma}_{A} \frac{C_{\bm{e}}}{V_0},
\end{align}
where $A$ is the area of the $ab$ plane.
\section{Computational Methods}
In order to test the uniaxial stress model in comparison to the uniaxial strain model, the axial CTEs of several materials were calculated \emph{ab initio} using both models. The selected materials (graphite, zinc, and calcite (Fig. \ref{cifs})) exemplify simple structures with highly anisotropic thermal and mechanical behaviour and their physical properties are well-known. \citep{bailey1970anisotropic, gauster1974pressure, meyerhoff1962anisotropic, ledbetter1977elastic, ramachandran1972generalised, rao1968precision, chen2001letters, dandekar1968temperature}  

Density functional theory calculations were carried out with the Abinit software package (v. 8.0.8) using pseudopotentials and plane-waves.  \citep{Gonze2016106, bottin2008large} All calculations were performed using the Perdew--Burke--Ernzerhof generalized gradient approximation to the exchange--correlation functional; \citep{perdew1996generalized} for graphite and calcite the vdw--DFT--D2 dispersion correction was added. \citep{grimme2006semiempirical} Optimized norm-conserving Vanderbilt pseudopotentials \citep{hamann2013optimized} from the Abinit library \cite{abipsps} were used in all cases; these pseudopotentials were tested by comparison of calculated elastic properties to experimental results. \cite{si} Plane-wave basis set energy cutoffs, Monkhorst--Pack grid spacings, \citep{monkhorst1976special} and van der Waals tolerance factors \citep{grimme2006semiempirical} were chosen through convergence studies.\cite{si} The values of these parameters can be found in tabular form in the Supplemental Material. \cite{si}

For each material, the structure was relaxed under conditions of zero external stress, and under uniaxial (biaxial) stress and strain perturbations along the $c$ axis ($ab$ plane). The magnitudes of the perturbations were generally chosen to give strains of 0.1\% for both the stress and strain cases. The phonon energies and elastic tensors of the relaxed geometries were calculated using DFPT;  \citep{gonze1997first, gonze1997dynamical, van2016interatomic} integration of phonon energies over the Brillouin zone yielded heat capacities. \citep{lee1995ab} \Gru parameters and axial CTEs were obtained from these data as described above. In the case of zinc, electronic contributions to the axial CTEs were included. \citep{barron1967analysis, verstraete2001smearing}
 
\section{Results}

\begin{figure}[htbp]
 \includegraphics[width=8cm]{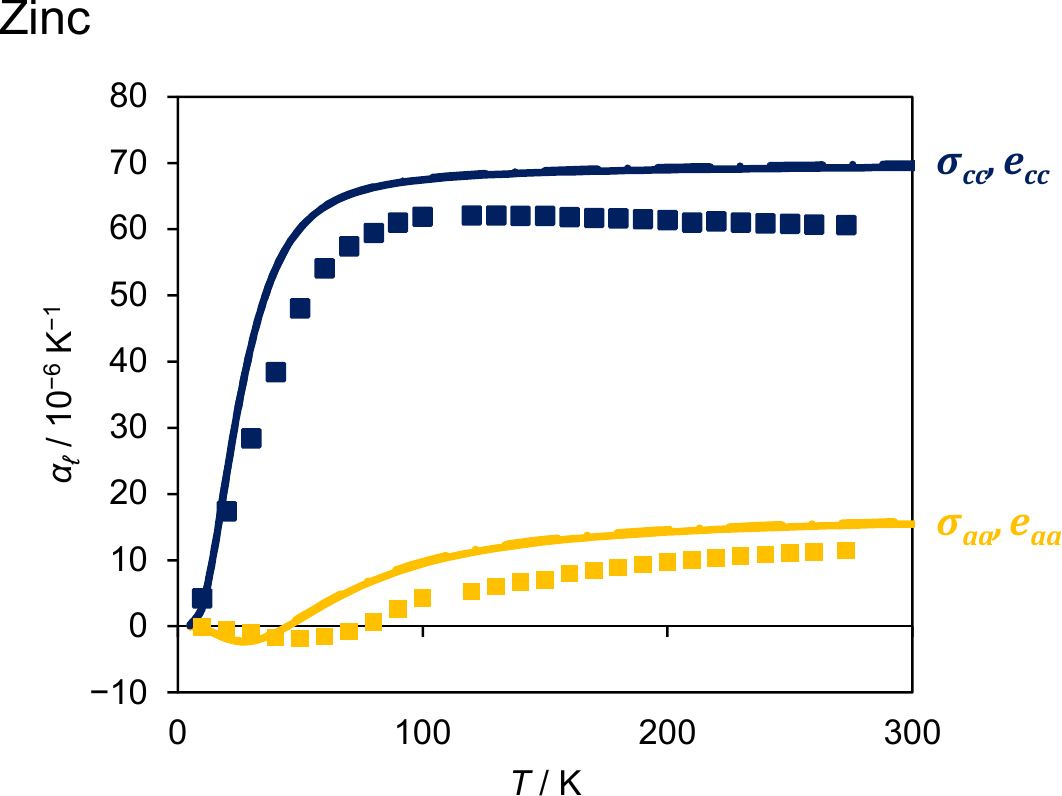}
 \includegraphics[width=8cm]{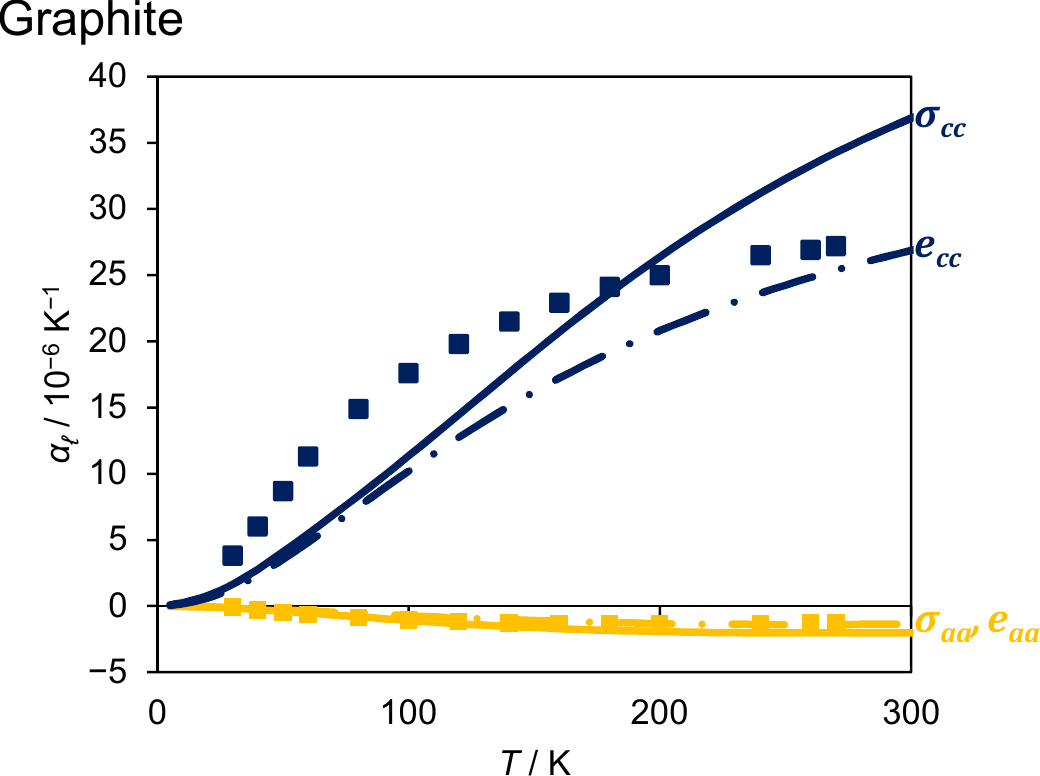}
 \caption{\label{znc}
 Linear thermal expansion in zinc and graphite along the $a$ (orange lines) and $c$ (blue lines) axes, as predicted by stress (solid lines) and strain (dashed lines) models. Experimental data \citep{bailey1970anisotropic,meyerhoff1962anisotropic} are shown as squares.}
\end{figure}

The first two materials considered, zinc and graphite, have very simple structures and similar thermoelastic properties. The stress and strain models used to predict their axial thermal expansion showed reasonable agreement with experimental data (Fig. \ref{znc}). The predicted $\alpha_{cc}$ in graphite was significantly lower than the experimental value at low temperature, despite the calculated phonon band structure and elastic tensor providing good matches to experiment (see Supplemental Material). \cite{si} However, the van der Waals nature of the interactions along $c$ provides a significant challenge for dispersion-corrected DFT. \citep{van2016interatomic,lechner2016first} Otherwise, the stress model of (Eq. (\ref{simple}))  produced identical results to that of the strain model (Eq. (\ref{bmalpha})).

\begin{figure}[htbp]
 \includegraphics[width=8cm]{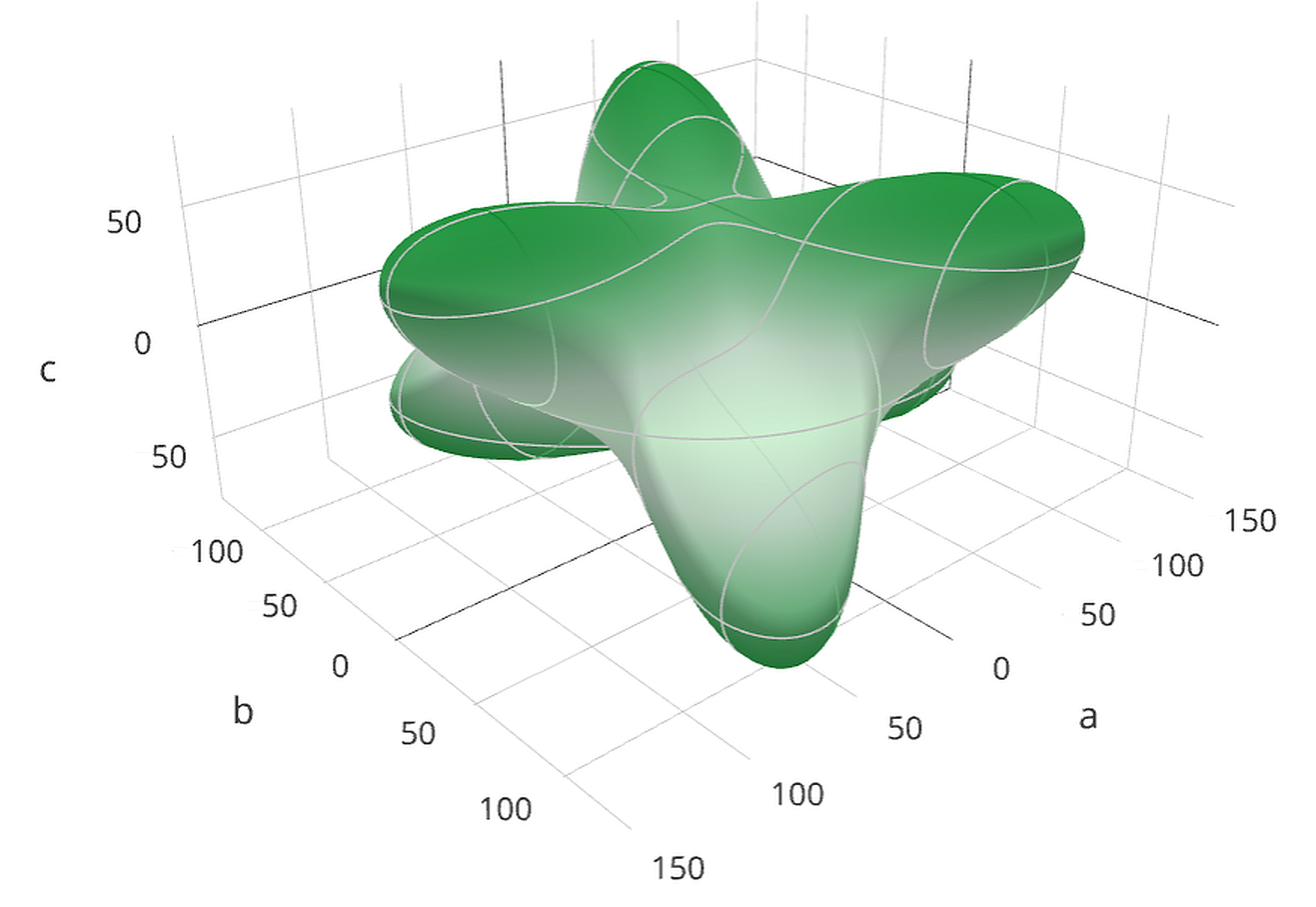}
 \includegraphics[width=8cm]{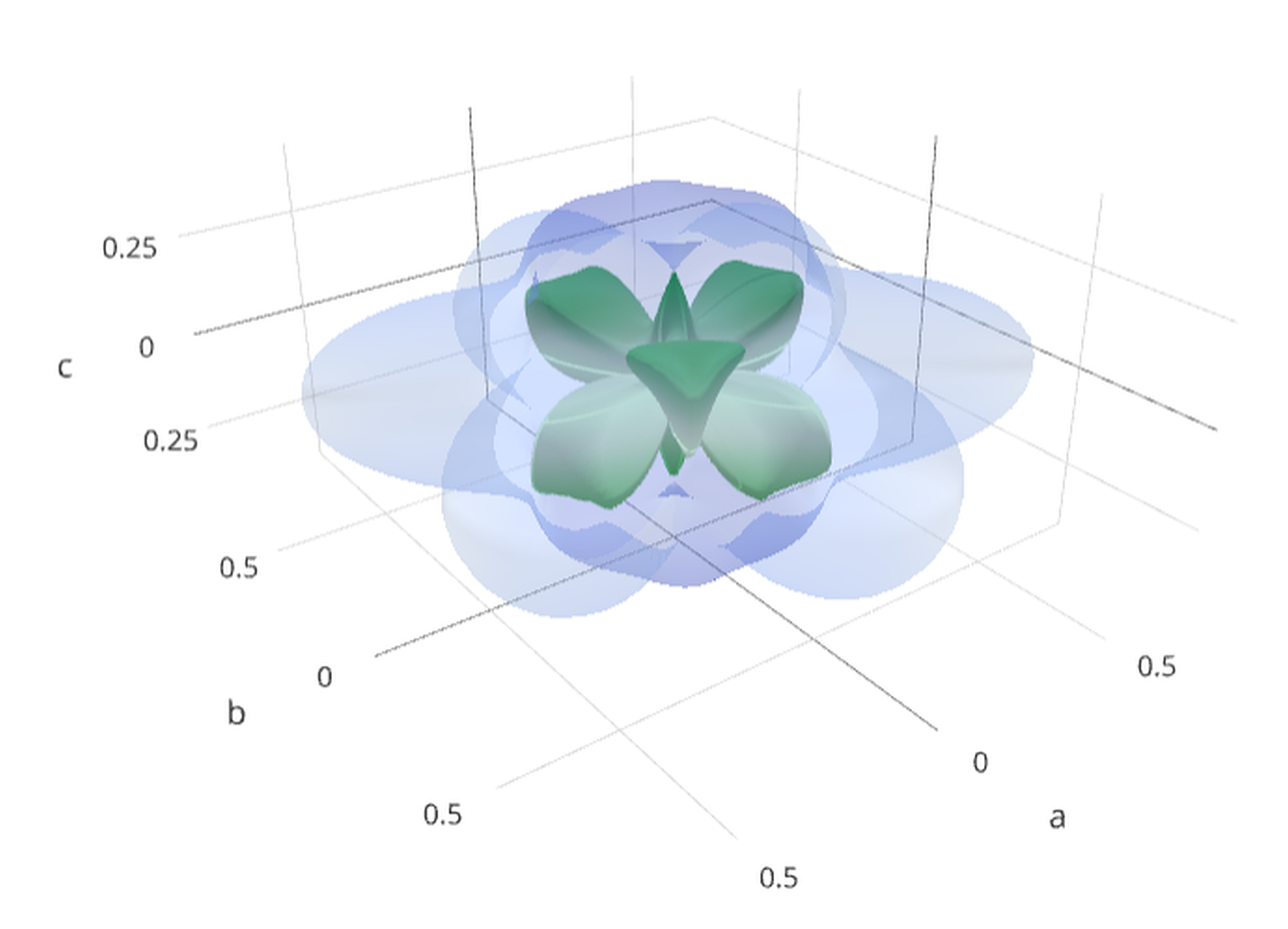}
 \caption{\label{caco3C}
Directional Young's moduli (in GPa, left) and Poisson ratios (right) of calcite. The Young's modulus in a given direction is shown as a green surface. The surface corresponding to the maximum Poisson ratio is shown in blue, and the surface corresponding to the minimum Poisson ratio is shown in green. Visualization generated with ELATE.  \cite{elate,gaillac2016elate}}
\end{figure}

Zinc and graphite have significant elastic anisotropy, as their $c$ axes are considerably more compliant than their $a$ axes, \citep{si,ledbetter1977elastic,gauster1974pressure} but the elastic couplings between the $a$ and $c$ axes are not unusually strong. (for zinc $\nu_{aacc} = 0.32$ and $\nu_{ccaa} = 0.13$; for graphite $\nu_{aacc} = -0.20$ and $\nu_{ccaa} = -0.008$). Since the stress and strain perturbations are identical in the limit of zero Poisson ratio, a more rigorous test can be obtained by considering a material with strong elastic couplings between axes. The calculated elastic tensor of calcite indicates that it has significant elastic couplings between its principal axes (Fig. \ref{caco3C}). The directional Young's moduli ($Y_{ii} = {s_{iiii}}^{-1}$) also show significant anisotropy (Fig. \ref{caco3C}), and therefore the elastic contribution to thermal expansion anisotropy in calcite is expected to be different from those of zinc and graphite.

\begin{figure}[htbp]
 \includegraphics[width=8cm]{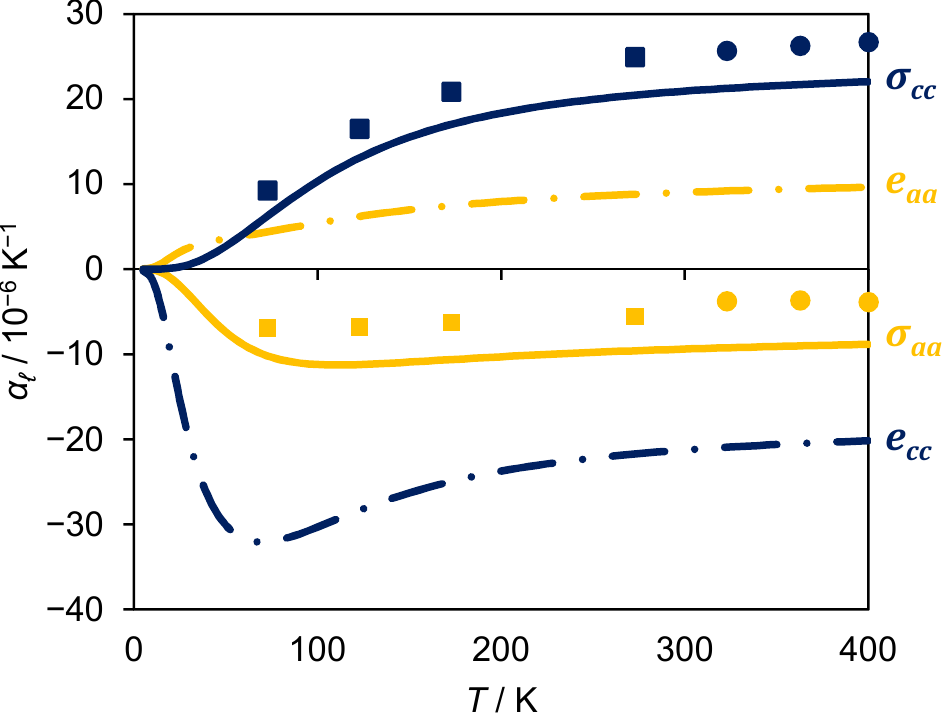}
 \caption{\label{caco3alpha}
Linear thermal expansion in calcite along the $a$ (orange lines) and $c$ (blue lines) axes, as predicted by stress (solid lines) and strain (dashed lines) models. Experimental data are shown as squares (Ref. \citenum{ramachandran1972generalised}) and circles (Ref. \citenum{rao1968precision}).}
\end{figure}

\begin{figure*}[htbp]
 \includegraphics[width=16cm]{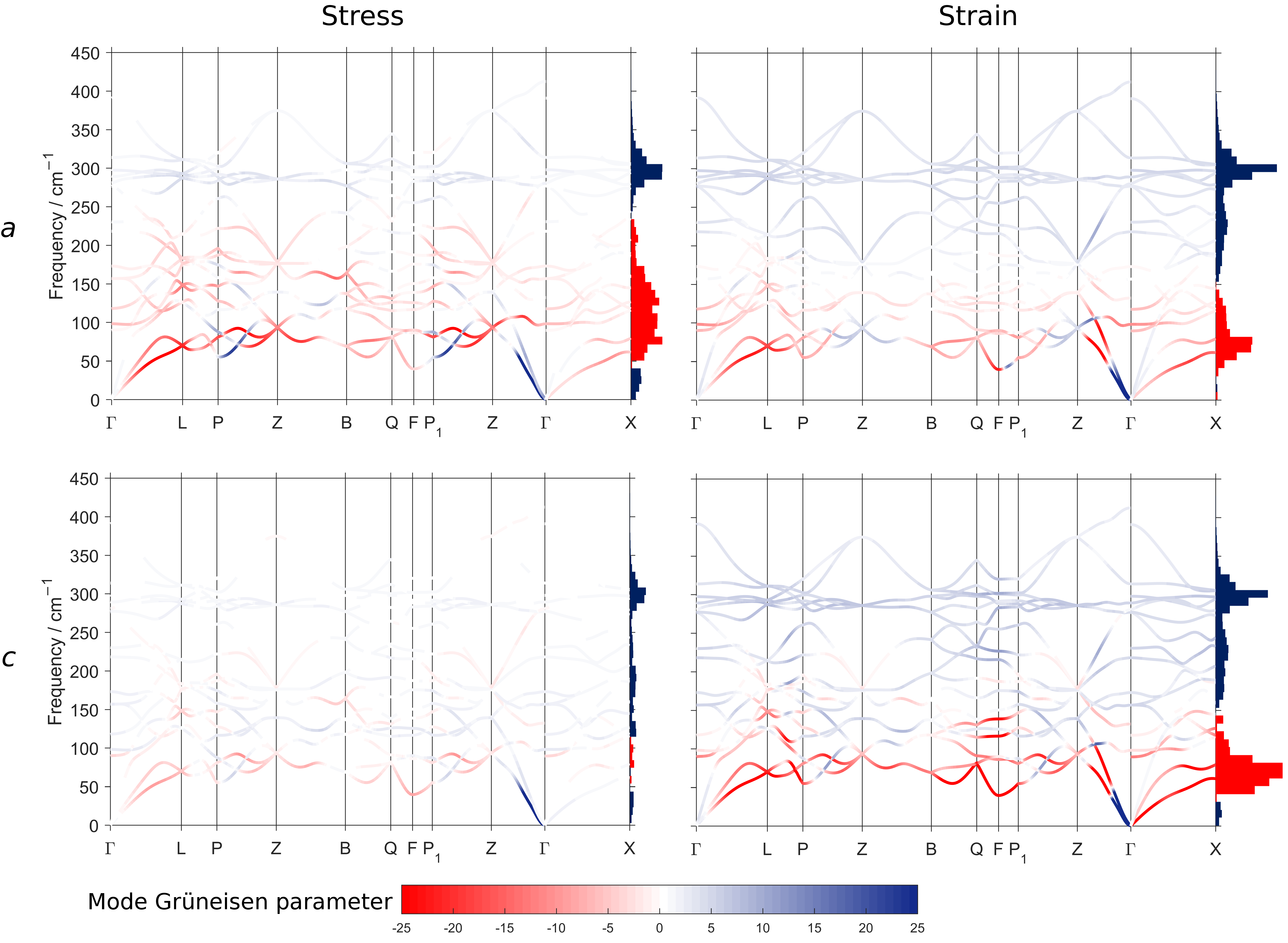}
 \caption{\label{caco3phonon}
Phonon band structure of calcite, with bands coloured according to their axial mode \Gru parameters calculated using stress and strain perturbations. Phonons with energies greater than 450 cm$^{-1}$ do not contribute significantly to thermal expansion and are not shown here. The density of states ($\rho$), weighted by the \Gru parameters as $\sum_{\mathbf{k}} \rho_{\mathbf{k}}(\omega) \gamma_{n,\mathbf{k}}(\omega)$, is shown as a histogram at the right of each plot, with positive values coloured in blue and negative values in red. Special points in and paths through the Brillouin zone were selected following Ref. \citenum{setyawan2010high}.}
\end{figure*}

Unlike in the cases of zinc and graphite, the stress and strain models gave significantly different predictions of axial thermal expansion in calcite (Fig. \ref{caco3alpha}); with the stress model providing a good match to the experimental data and the strain model erroneously predicting $\alpha_{aa}$ to be positive and $\alpha_{cc}$ to be negative. Thermal expansion anisotropy in calcite is driven by low-energy acoustic and optic modes (Fig. \ref{caco3phonon}) in which the CO$_{3}^{2-}$ unit remains rigid. The acoustic modes which propagate along $c$ (with wavevector $\Gamma$--Z) have large positive \Gru parameters with respect to all perturbations, while those which propagate in the $ab$ plane have negative \Gru parameters. The group of optic modes with negative mode \Gru parameters below 150 cm$^{-1}$ involves librations of the CO$_{3}^{2-}$ unit, while the group between 150 and 450 cm$^{-1}$ includes motion of the Ca$^{2+}$ ion, although there is considerable eigenvector mixing away from $\Gamma$.\cite{si} This view of calcite as, in some respects, a framework solid, is supported by the directional Young's moduli showing maxima coinciding with the directions of Ca--O--C linkages and by the large Poisson ratios in these directions (Fig. \ref{caco3C}).

The \Gru parameters obtained from the stress perturbations indicate that negative thermal expansion along $a$ is driven by low-energy acoustic and librational modes. Along $c$, the \Gru parameters are small and mostly positive; the reduced stiffness along $c$ increases $\alpha_{cc}$. By contrast, the mode \Gru parameters related to the strain perturbation along $a$ and along $c$ are similar. Due to the significant Poisson ratios relating $a$ and $c$ ($\nu_{aacc} = 0.45$ and $\nu_{ccaa} = 0.26$) the stresses transverse to the strain perturbation are of the same order of magnitude as the stresses along the perturbation direction. Therefore, the inaccuracy of the uniaxial strain model in this case indicates that the convolution of the axial \Gru parameters through the cross-compliances (Eq. (\ref{bmalpha})) is inexact.

\section{Discussion}
\begin{figure}[htbp]
 \includegraphics[width=8cm]{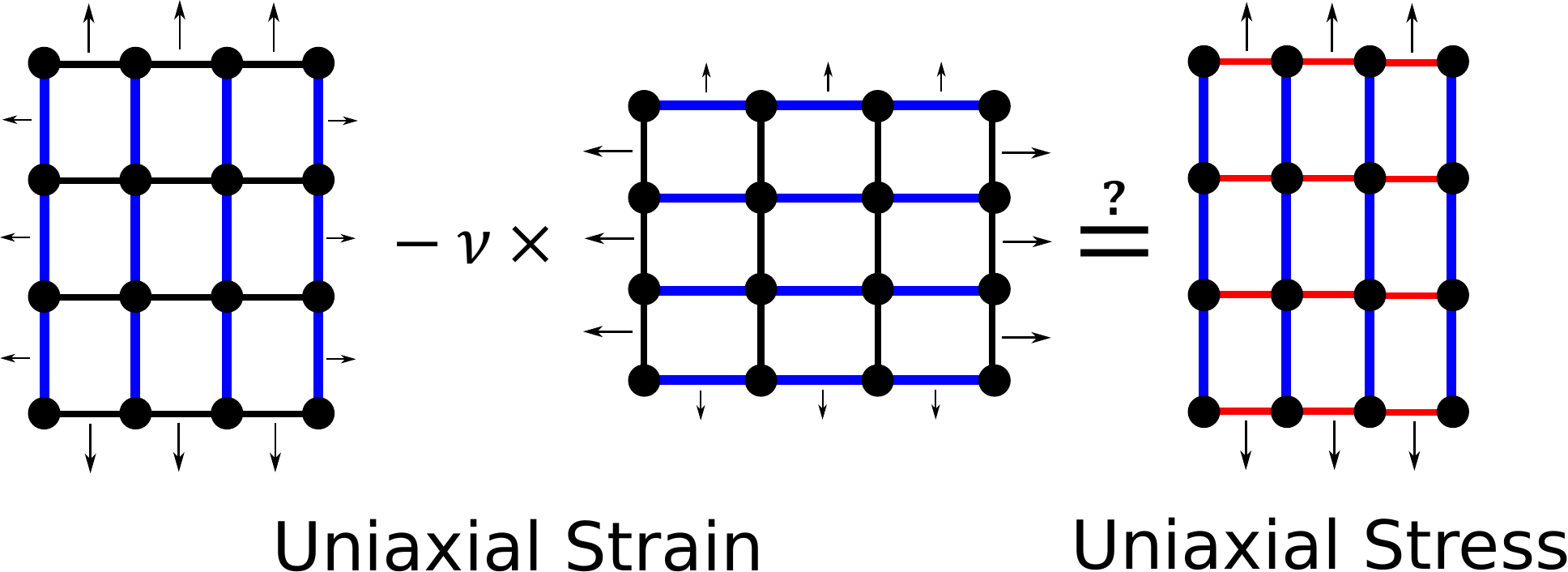}
 \caption{\label{compare}
 A comparison of the uniaxial strain and stress models for a simplified system with positive thermal expansion and positive Poisson ratio. Bonds coloured in blue are lengthened by the perturbation, leading to a decrease in vibrational frequency and a positive contribution to the bulk \Gru parameter, while those coloured in red are contracted by the Poisson effect, increasing their frequency and therefore reducing the \Gru parameter.}
 \end{figure}

The similarities and differences between the uniaxial stress perturbation (Eq. (\ref{simple})) and the uniaxial strain perturbation (Eq. (\ref{bmalpha})) can be appreciated by considering their application to a simplified model. Fig. \ref{compare} shows a square lattice with a positive Poisson ratio and positive thermal expansion, where each bond vibrates independently. When the lattice is subjected to a uniaxial strain perturbation, the bonds aligned with the perturbation elongate and their vibrational frequencies decrease, indicating a positive contribution to $\alpha$. However, a negative contribution to $\alpha$ comes from the bonds orthogonal to the perturbation, proportional to the Poisson ratio relating the two axes. In the uniaxial stress case, the Poisson effect contracts the bonds perpendicular to the perturbation, again resulting in a decrease in $\alpha$ proportional to the Poisson ratio. It can therefore be appreciated that, for this simplified model, Eq. (\ref{bmalpha}) and Eq. (\ref{simple}) are equivalent.

In the simplified model, the vibrational frequencies are linearly related to the lattice constants. This requires two predicates: that the vibrational frequencies are proportional to interatomic distances, and that the interatomic distances are proportional to the lattice constants. The first is a form of the QHA, stating that phonon energies can be expressed as a function of internal strain coordinates. \citep{hofmeister2002redefinition} The second is geometric: in the simplified model, there are no atomic coordinates which are not fixed by the lattice. If this is not the case, the bond lengths will not, in general, scale linearly with the lattice vectors, and the stress and strain models will be inequivalent. This can occur if the relative positions of the atoms are not fixed by symmetry.


Therefore, the differences between the stress model and the strain model for the materials studied herein (Fig. \ref{znc} and Fig. \ref{caco3alpha}) can be explained by their structures. The atomic coordinates of zinc and graphite are fixed by the lattice constants, and therefore are analogous to the simple structure of Fig. \ref{compare}, and the stress and strain models give results of comparable accuracy. Unlike zinc and graphite, calcite features an internal coordinate not fixed by the lattice constants, and flexible Ca--O--C linkages. This, in combination with the large Poisson ratios in calcite (Fig. \ref{caco3C}), leads to the large discrepancy between the two models seen in Fig. \ref{caco3alpha}.

The increased accuracy of the stress model relative to the strain model seen in the \emph{ab initio} calculations of $\bm{\alpha}$ presented herein can therefore be attributed to the assumption of the strain model that thermal strains along different axes are coupled purely elastically. This treatment ignores that the internal strain coordinates relevant to a particular mode may not have the same elastic behaviour as the lattice. When performing a uniaxial stress perturbation, the Poisson effect is included directly in the model, and no correction for the transverse stresses is required. Since the magnitude of this correction is determined by the cross-compliances, for many systems the difference between the two models is relatively small. However, it will be especially important for materials with unusual elastic properties.


The uniaxial stress model also offers other advantages to the understanding of the origins of thermal expansion. Coupling between thermal expansion and elasticity can be understood in a simpler way, as the \Gru parameter along one axis and one element of the compliance tensor determine the CTE in that direction without reference to the transverse axes. Therefore, negative thermal expansion is impossible without modes with negative \Gru parameters. In fact, although the strain model allows for negative thermal expansion from positive \Gru parameters due to the Poisson effect, the only materials where it has been suggested that this occurs are zinc and cadmium. \citep{barrera2005negative} 

The appearance of $s_{ijij}$ in Eq. (\ref{simple}) indicates that directional thermal expansion can be predicted by reference to the directional Young's moduli ($Y_{ii} = {s_{iiii}}^{-1}$). This was perhaps anticipated by Barker Jr., \citep{barker1963approximate} who found that for a broad range of materials the approximate relationship $Y\alpha^2 \approx 15~ \text{Pa}$ holds, and that differences in thermal expansivity between materials are often driven by their relative Young's moduli rather than by differences in the \Gru parameter. This approach can be extended by considering, for example, directional Young's moduli in calcite (Fig. \ref{caco3C}) in relation to directional thermal expansion. Fig. \ref{caco3C} and Fig. \ref{cifs} show that the calcite structure is most stiff along directions  corresponding to Ca--O--C linkages. Rotation of $\bm{\alpha}$ shows that the directions of maximum stiffness have very low thermal expansion ($\alpha_{\ell} = 6 \times 10^{-8}~\text{K}^{-1}$). In fact, if by inspection of Fig. \ref{caco3C} one was to assume that the stiffest directions have smaller magnitudes of $\alpha_{\ell}$ than do the principal axes, this would lead to the conclusion that $\alpha_{\ell}$ must be negative along one principal axis and positive along the other, based on the required symmetry of $\bm{\alpha}$ (\emph{i.e.}, that the maxima and minima lie along principal axes).\citep{nye1985physical} 

This analysis can be extended to other orthotropic systems where stiffness maxima are not aligned with the unit cell vectors, \emph{e.g.} the metal-organic wine-rack framework material MIL-53 has $\alpha_{\ell} = 7 \times 10^{-7}~\text{K}^{-1}$ along the stiff wine-rack axes, leading to anomalous thermal expansion along the compliant principal axes ($\alpha_{bb} = -1.4 \times 10^{-5}~\text{K}^{-1}$, $\alpha_{cc} = 2.4 \times 10^{-5}~\text{K}^{-1}$).\citep{nanthamathee2014contradistinct} When the \Gru parameter along the stiffest direction is anomalous, even more unusual behaviour can occur. For example, Ag$_3$Co(CN)$_6$ has $\alpha_{\ell} = -2.5 \times 10^{-5}~\text{K}^{-1}$ along a Co--CN linkage (a typical value for an M--CN chain);\citep{chippindale2012mixed} this, along with the compliance of the $ab$ plane, results in colossal positive and negative thermal expansion along the principal axes ($\alpha_{aa} = 1.4 \times 10^{-4}~\text{K}^{-1}$, $\alpha_{cc} = -1.3 \times 10^{-4}~\text{K}^{-1}$).\citep{goodwin2008colossal} This misalignment mechanism can be expected to occur commonly in materials which exhibit negative linear compressibility, which requires a mixture of stiff and compliant directions to balance stability and flexibility.\citep{cairns2015negative} Of course, the phenomenon is essentially geometric, and coincides with the geometric arguments previously used to explain anomalous thermal expansion in these materials.\citep{goodwin2008colossal, nanthamathee2014contradistinct} However, removing the cross-coupling term of the strain model facilitates understanding of relationships between thermal expansion anisotropy and framework flexibility by removing the need to consider the (often large) Poisson ratios directly.

The stress model has an additional advantage over the strain model in that one element of $\bm{\alpha}$ can be calculated independently of the others. This offers the possibility of, for example, calculating one element in order to understand the mechanisms of uniaxial negative thermal expansion,\citep{senn2015negative} or to test the accuracy of an exchange-correlation functional or a set of pseudopotentials for a given system. Especially for monoclinic and triclinic crystal families, the computational expense required to calculate \Gru parameters for every element of $\bm{\alpha}$ may be prohibitive, but a qualitative understanding of thermoelastic behaviour could perhaps be obtained with some subset thereof.

\section{Conclusions}
A \Gru model for anisotropic materials based on uniaxial strain perturbations has been proposed. This model has the advantage of including the mechanical coupling between axes explicitly, allowing  the thermal expansion axis to be related to mode \Gru parameters and the Young's modulus in that direction only. The model was tested by \emph{ab initio} prediction of thermal expansion in several highly anisotropic materials; revealing that the uniaxial stress model has equal or better accuracy to the previous uniaxial strain model. By relating the directional Young's moduli to thermal expansion directly, it can be predicted that framework materials whose rigid units are misaligned with the principal axes are likely to display positive and negative axial thermal expansion.

\begin{acknowledgments}
This study was supported by the Natural Sciences and Engineering Research Council of Canada (NSERC) and the University of Oxford, Department of Chemistry.
\end{acknowledgments}
\bibliography{gruneisen}
\bibliographystyle{apsrev}

\end{document}